\documentclass{article}
\usepackage{spconf,amsmath,graphicx}
\usepackage{amsfonts}
\usepackage{amssymb}
\usepackage{algorithmic}
\usepackage{algorithm}
\usepackage{color, soul}
\usepackage{stfloats}
\usepackage{setspace}
\usepackage{algorithm}
\usepackage{algorithmic}
\usepackage{float}
\usepackage{verbatim}
\usepackage{cite}
\usepackage{url}
\usepackage{bm}
\usepackage{stmaryrd}
\usepackage{multicol}
\usepackage{stfloats}
\usepackage[amsmath,thmmarks]{ntheorem}
\usepackage{theorem}
\usepackage[justification=centering]{caption}

\theoremheaderfont{\sc}\theorembodyfont{\upshape}
\theoremstyle{nonumberplain}
\theoremseparator{}
\theoremsymbol{\rule{1ex}{1ex}}

\usepackage{hyperref}
\title{A Unitary Transform Based Generalized Approximate Message Passing}
\name{Jiang Zhu$^{\ddag}$ \qquad Xiangming Meng$^{\star}$\thanks{Corresponding author. E-mail: meng@g.ecc.u-tokyo.ac.jp. This work is supported by NSFC under grant 61901415.} \qquad Xupeng Lei$^{\ddag}$\qquad Qinghua Guo$^{\dagger}$}

\address{$^{\ddag}$ Ocean College, Zhejiang University, Zhoushan, 316021, China \\
$^{\star}$Institute for Physics of Intelligence,
The University of Tokyo, 7-3-1, Hongo, Tokyo 113-0033, Japan\\
$^{\dagger}$SECTE, University of Wollongong, NSW, 2522, Australia}

\begin{document}
\maketitle
\begin{abstract}
\vspace{-0.5em}
We consider the problem of recovering an  unknown signal ${\mathbf x}\in {\mathbb R}^n$ from general nonlinear measurements obtained through a generalized linear model (GLM), i.e., ${\mathbf y}= f\left({\mathbf A}{\mathbf x}+{\mathbf w}\right)$, where $f(\cdot)$ is a componentwise nonlinear function. Based on the unitary transform approximate message passing (UAMP) and  expectation propagation,  a unitary transform based generalized approximate message passing (GUAMP) algorithm is proposed for general measurement matrices $\bf{A}$, in particular highly correlated matrices. Experimental results on quantized compressed sensing demonstrate that the proposed GUAMP significantly outperforms  state-of-the-art GAMP and GVAMP under correlated matrices $\bf{A}$.
\end{abstract}
\vspace{-1em}
\section{Introduction}
\vspace{-0.6em}
We consider the general problem of inference on generalized linear models (GLM) \cite{mccullagh2019generalized}, i.e., recovering an unknown signal ${\mathbf x}\in {\mathbb R}^n$  from a noisy linear transform followed by componentwise nonlinear measurements
\vspace{-0.5em}
\begin{align}
    {\mathbf y}= f\left({\mathbf A}{\mathbf x}+{\mathbf w}\right), \label{eq:GLM-def}
\end{align}
where ${\bf{A}} \in \mathbb{R} ^{m \times n}$ is a known linear mixing matrix,  ${\bf{w}}\sim \mathcal{N}({\bf{w}};0,\sigma^2 {\bf{I}}_m)$  is an i.i.d.  Gaussian noise with known variance, and ${f}(\cdot): \mathbb{R}^{m \times 1} \to \mathcal{Q}^{m \times 1} $ is an \textit{componentwise} nonlinear link function. Denote ${\mathbf z}\triangleq {\mathbf A}{\mathbf x}\in{\mathbb R}^m$ as the hidden linear transform outputs, the componentwise nonlinear  function $f(\cdot)$ can be equivalently described in a probabilistic way using a fully factorized likelihood function as follows
\vspace{-0.55em}
\begin{align}
p({\mathbf y}|{\mathbf z})=\prod\limits_{i=1}^mp(y_i|z_i),
\end{align}
where $p(y_i|z_i)$ is determined by the specific nonlinear function $f(\cdot)$. The prior distribution $p({\mathbf x})$ of $\bf{x}$ is also assumed to be fully factorized $p({\mathbf x})=\prod\limits_{j=1}^np(x_j)$ for simplicity.  GLM inference has wide applications in science and engineering such as  wireless communications, signal processing, and machine learning. In the special case of identity function $f(\cdot)$, the nonlinear GLM in (\ref{eq:GLM-def}) will reduce to the popular standard linear model (SLM) as follows
\vspace{-0.55em}
\begin{equation}\label{SLM}
{\mathbf y}={\mathbf A}{\mathbf x}+{\mathbf w}.
\end{equation}
Thus, GLM is actually an extention of SLM from linear measurements to nonlinear measurements, which are prevalent in some real-world applications such as quantized compressed sensing, pattern classification, phase retrieval, etc.

A variety of algorithms have been proposed for inference over GLMs and SLMs. Among them, the past decade has witnessed an advent of one distinguished family of probabilistic algorithms called  message passing algorithm. Among them,  the most famous ones are the approximate message passing (AMP) algorithm \cite{kabashima2003cdma,donoho2009message} for SLM and generalized approximate message passing (GAMP) \cite{Rangan1} for GLM, which have been proved to be optimal under i.i.d. Gaussian matrices $\bf{A}$. However, both AMP and GAMP diverge for general $\bf{A}$. In \cite{GAMPSBLTSP, Dirk1, Dirk2}, the GAMP is incorporated into the sparse Bayesian learning (SBL) to improve the robustness of GAMP and reduce the computation complexity of SBL. Vector AMP (VAMP) \cite{VAMP} (or OAMP \cite{OAMP}, one similar algorithm to VAMP) and generalized VAMP (GVAMP) \cite{GVAMP} have been proposed to improve the performance with general $\bf{A}$ for SLM and GLM, respectively. The VAMP is derived from expectation propagation (EP) \cite{Minka}, another powerful message passing algorithm. Remarkably, it has been
demonstrated in \cite{mengEP,meng2} that all the AMP, GAMP, and GVAMP, can be derived concisely as special instances of EP under different assumptions and thus be unified within a single EP framework.  Despite significant improvement in robustness, in particular for right-orthogonally invariant  matrices,  VAMP and GVAMP still suffer poor convergence in some more challenging scenarios, e.g., the measurement matrix is highly correlated \cite{ivrlac2003fading,chizhik2003multiple}.

In this work, we focus on the AMP variant: unitary approximate message passing (UAMP) \cite{Guo}, which was formerly called UTAMP \cite{Guo, GuoUAMPSBL, UAMPSBL}. The motivation behind UAMP is this: since AMP has proven to work well when the elements of $\mathbf A$ in (\ref{SLM}) are uncorrelated, one can artificially constructs such a ``good" and equivalent measurement model simply by  performing a  singular value decomposition (SVD) on ${\mathbf A}={\mathbf U}{\boldsymbol \Sigma}{\mathbf V}^{\rm T}$, where ${\mathbf U}\in{\mathbb R}^{m\times r}$, ${\boldsymbol\Sigma}\in{\mathbb R}^{r\times r}$, ${\mathbf V}\in{\mathbb R}^{n\times r}$ and $r={\rm rank}({\mathbf A})$, and by left multiplying ${\mathbf U}^{\rm T}$ on the original SLM (\ref{SLM}), thus leading to an equivalent SLM as \cite{Guo}
\vspace{-0.5em}
\begin{align}\label{UTSLM}
\tilde{\mathbf b}={\mathbf Q}{\mathbf x}+\tilde{\mathbf w},
\end{align}
where $\tilde{\mathbf b}={\mathbf U}^{\rm T}{\mathbf y}\in{\mathbb R}^r$, ${\mathbf Q}={\boldsymbol \Sigma}{\mathbf V}^{\rm T}\in{\mathbb R}^{r\times n}$ and $\tilde{\mathbf w}={\mathbf U}^{\rm T}{\mathbf w}\in{\mathbb R}^{r}$ correspond to unitary-transformed pseudo linear measurements, linear mixing matrix, and additive noise, respectively. Subsequently, one can readily run standard AMP on the equivalent model (\ref{UTSLM}), which leads to one version of UAMP. Further, two averaging operations can be performed to obtain the other version of UAMP, where the number of matrix-vector products per iteration is reduced from 4 to 2 \cite{Guo, GuoUAMPSBL, UAMPSBL}. Despite its simplicity, UAMP has been shown to be more robust than AMP and VAMP under some types of ``tough" measurement matrices, e.g., highly correlated matrices \cite{GuoUAMPSBL, UAMPSBL}. Nevertheless, the extension of UAMP to the case of GLM has remained lacking.

In this paper, we extend UAMP from SLM to GLM and propose the generalized UAMP (GUAMP). The key idea is to utilize the unified Bayesian inference framework in \cite{meng2,zhucomment} and EP \cite{Minka} to iteratively  decompose the original nonlinear measurement model (\ref{eq:GLM-def}) into a series of pseudo SLMs, whereby UAMP could be conducted. Conceptually, GUAMP consists of
two modules, namely AMP module and GAMP module, as shown in Fig. \ref{Fg_Module}. Extrinsic information \cite{Exdef} is exchanged per iteration between the two modules before  GUAMP finally converge. Experimental results demonstrate that GUAMP significantly outperforms both GAMP and GVAMP under highly correlated matrices $\bf{A}$.

\vspace{-0.8em}
\section{The GUAMP Algorithm}
\vspace{-1em}
The key idea is that, inspired by UAMP\cite{Guo}, we introduce an additional hidden variable ${\mathbf b}\in{\mathbb R}^{r}$ and thus an equivalent representation of ${\mathbf z}\triangleq {\mathbf A}{\mathbf x}$ as follows
\vspace{-0.5em}
\begin{subequations}
\begin{align}
&{\mathbf b}={\boldsymbol\Sigma}{\mathbf V}^{\rm T}{\mathbf x}\triangleq {\mathbf Q}{\mathbf x},\\
&{\mathbf z}={\mathbf U}{\mathbf b},
\end{align}
\end{subequations}
where, as in the original UAMP \cite{Guo},  ${\mathbf U}\in{\mathbb R}^{m\times r}$, ${\boldsymbol\Sigma}\in{\mathbb R}^{r\times r}$, ${\mathbf V}\in{\mathbb R}^{n\times r}$ are obtained from SVD ${\mathbf A}={\mathbf U}{\boldsymbol \Sigma}{\mathbf V}^{\rm T}$ when ${\rm rank}({\mathbf A})=r$. Note that the underdetermined ($m\leq n$) and overdetermined ($m\leq n$) cases are treated  in a unified way.

\begin{figure}[t!]
		\centering
		\includegraphics[width=3.0in]{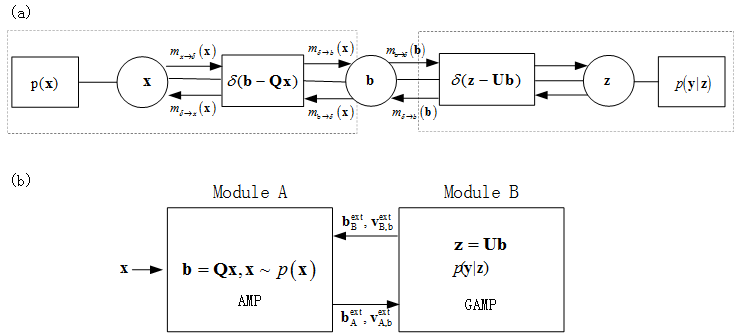}
		\caption{(a). Equivalent factor graph representation of GLM. (b) Modular representation of the GUAMP algorithm. }
		\label{Fg_Module}
\end{figure}

The corresponding factor graph of the original GLM (\ref{eq:GLM-def}) can be equivalently shown  in Fig.\ref{Fg_Module} (a). Subsequently, using the unified modular framework in \cite{meng2}, the inference on the factor graph in Fig.\ref{Fg_Module} (a) can be decomposed into two modules, namely one AMP module and one GAMP module, as shown in Fig.\ref{Fg_Module} (b). At a high level, in each iteration, module A performs standard AMP w.r.t. $\bf{x}$ on an equivalent SLM with a pseudo measurement matrix $\bf{Q}$ while module B performs GAMP w.r.t. $\bf{b}$  on an equivalent GLM with a pseudo measurement matrix $\bf{U}$, rather than  the original measurement matrix $\bf{A}$. Module A and Module B exchange extrinsic information with each other in the same way as \cite{meng2} and this process proceeds before convergence. Please refer to \cite{meng2} for further details of the modular representation perspective via EP \cite{Minka}.

In the following, we describe the implementation details of GUAMP. First of all, we initialize ${\boldsymbol\tau}^b(0)\in{\mathbb R}^r$, $\hat{\mathbf s}_{\rm B}(-1)\in {\mathbb R}^m$, ${\mathbf b}_{{\rm A}}^{\rm ext}(1)\in {\mathbb R}^r$ and variances ${\mathbf v}_{{\rm A}}^{{\rm ext},b}(1)\in {\mathbb R}^r$ in module B, and initialize $\hat{\mathbf p}_{\rm A}(0)\in {\mathbb R}^r$,  $\hat{\boldsymbol\tau}_{\rm A}^p(0)\in {\mathbb R}^r$, $\hat{\mathbf x}(0)\in {\mathbb R}^n$. The number of outer iterations between module A and module B is set as  $\rm T_{\rm max}$, while the number of inner iterations  in module A and B during an outer iteration are set as $T_{\rm A}$ and $T_{\rm B}$, respectively. Next, we present the details of the operation in the two modules.
\subsection{The GAMP Module}
\vspace{-0.5em}
The extrinsic means  ${\mathbf b}_{{\rm A}}^{\rm ext}(t)\in {\mathbb R}^r$ and variances ${\mathbf v}_{{\rm A}}^{{\rm ext},b}(t)\in {\mathbb R}^r$ transmitting from module A to module B can be viewed as the prior means and variances of $\mathbf b$. i.e.,
\begin{align}
p({\mathbf b})={\mathcal {N}}({\mathbf b};{\mathbf b}_{{\rm A}}^{\rm ext}(t),{\rm diag}({\mathbf v}_{{\rm A}}^{{\rm ext},b}(t))).
\end{align}
In addition, ${\mathbf z}={\mathbf U}{\mathbf b}$ and ${\mathbf y}|{\mathbf z}\sim p({\mathbf y}|{\mathbf z})$. Thus we could run the GAMP algorithm treating $\bf{b}$ as the unknown signal and $\bf{U}$ as the measurement matrix as follows.
\begin{itemize}
\item\quad [Step 1 (B)] Perform output linear step to obtain $\hat{\mathbf p}_{{\rm B}}(t)\in{\mathbb R}^m$ and ${\boldsymbol \tau}^p_{{\rm B}}(t)\in{\mathbb R}^m$ as
\begin{subequations}
		\begin{align}
			{\boldsymbol\tau}^p_{{\rm B}}(t)&=|{\mathbf U}|^2{\boldsymbol \tau}^b(t), \label{taupBvec}\\
			\hat{\boldsymbol p}_{{\rm B}}(t)&={\mathbf U}\hat{\mathbf b}(t)-{\boldsymbol \tau}^p_{{\rm B}}(t)\hat{\mathbf s}_{{\rm B}}(t-1). \label{hatpBvec}
		\end{align}
\end{subequations}
  \item \quad[Step 2 (B)] Perform output linear step to obtain $\hat{s}_{{\rm B},i}(t)$ and ${\tau}_{{\rm B},i}(t)$ as
\begin{subequations}
\begin{align}
\hat{s}_{{\rm B},i}(t)=g_{\rm out}(\hat{p}_{{\rm B},i}(t),y_i,\tau^p_{{\rm B},i}(t)),\label{hatsB}\\
{\tau}_{{\rm B},i}(t)=-\frac{\partial g_{\rm out}(\hat{p}_{{\rm B},i}(t),y_i,\tau^p_{{\rm B},i}(t))}{\partial \hat{p}}, \label{tauB}
\end{align}
\end{subequations}
for $i=1,2,\cdots,m$, where
\begin{align}
g_{\rm out}(\hat{p},y,\tau^p)&=\frac{\hat{z}^{0}-\hat{p}}{\tau_p},\\
-\frac{\partial g_{\rm out}(\hat{p},y,\tau^p)}{\partial \hat{p}}&=\frac{\tau_p-{\rm Var}(z|\hat{p},y)}{(\tau_p)^2},
\end{align}
and the posterior means $\hat{z}^{0}$ and variances ${\rm Var}(z|\hat{p},y)$ are computed w.r.t. the posterior $\propto {\mathcal {CN}}(z;\hat{p},\tau_p)p(y|z)$. See \cite{Rangan1} for further details.
\item \quad[Step 3 (B)] Perform input linear step as
	\begin{subequations}
		\begin{align}
		{\boldsymbol\tau}^r_{{\rm B}}(t)&=\left( |{\mathbf U}^{\rm T}|^2{\boldsymbol\tau}^s_{{\rm B}}(t)\right)^{-1},\label{taurB}\\
		\hat{\mathbf r}_{{\rm B}}(t)&=\hat{\mathbf b}(t)+{\boldsymbol\tau}^r_{{\rm B}}(t)\odot \left({\mathbf U}^{\rm T}\hat{\mathbf s}_{{\rm B}}(t-1)\right).\label{hatrB}
		\end{align}
	\end{subequations}
  \item \quad[Step 4 (B)] Perform input nonlinear step in Module B to obtain the posterior means and variances of variable $\mathbf{b}$ as: For $j=1,2,\cdots,r$, 
	\begin{subequations}
		\begin{align}
			\hat{b}_j(t)&=\frac{\hat{r}_{{\rm B},j}(t){v}_{{\rm A},j}^{{\rm ext},b}(t)+{b}_{{\rm A},j}^{\rm ext}(t)\tau^r_{{\rm B},j}(t)}{\tau^r_{{\rm B},j}(t)+{v}_{{\rm A},j}^{{\rm ext},b}(t)}, \label{hatb}\\
			\tau_j^b(t)&=\frac{{v}_{{\rm A},j}^{{\rm ext},b}(t)\tau^r_{{\rm B},j}(t)}{\tau^r_{{\rm B},j}(t)+{v}_{{\rm A},j}^{{\rm ext},b}(t)}. \label{taub}
		\end{align}
	\end{subequations}
\end{itemize}
After running $T_{\rm B}$ iterations, the extrinsic means ${\mathbf b}_{{\rm B}}^{\rm ext}(t)$ and variances ${\mathbf v}_{{\rm B}}^{{\rm ext},b}(t)$ of $\mathbf b$ from module B to module A can be calculated as follows \cite{meng2}
\begin{align}
{\mathbf b}_{{\rm B}}^{\rm ext}(t)=\hat{\mathbf r}_{{\rm B}}(t),\\
{\mathbf v}_{{\rm B}}^{{\rm ext},b}(t)={\boldsymbol\tau}^r_{{\rm B}}(t).
\end{align}
\textit{Remark}: It is worth noting that if the GLM (\ref{eq:GLM-def}) degenerates to  the SLM (\ref{SLM}), it can be verified that the extrinsic information $\bf{b}_{{\rm B}}^{\rm ext}$ and ${\mathbf v}_{{\rm B}}^{{\rm ext},b}(t)$ from  module B to module A are always ${\mathbf b}_{{\rm B}}^{\rm ext}={\mathbf U}^{\rm T}{\mathbf y}$ and ${\mathbf v}_{{\rm B}}^{{\rm ext},b}=\sigma^2{\mathbf 1}_r$. Consequently, the GUAMP reduces to the UAMP in the special case of SLM.

\subsection{The AMP Module}	
\begin{algorithm}[ht]
\caption{GUAMP Algorithm}\label{GUAMP}
\begin{algorithmic}[1]
\STATE Initialize ${\boldsymbol\tau}^b(0)\in{\mathbb R}^r$, $\hat{\mathbf s}_{\rm B}\in {\mathbb R}^m$, ${\mathbf b}_{{\rm A}}^{\rm ext}(1)\in {\mathbb R}^r$ and variances ${\mathbf v}_{{\rm A}}^{{\rm ext},b}(1)\in {\mathbb R}^r$ in module B, initialize $\hat{\mathbf p}_{\rm A}(0)\in {\mathbb R}^r$,  $\hat{\boldsymbol\tau}_{\rm A}^p(0)\in {\mathbb R}^r$, $\hat{\mathbf x}(0)\in {\mathbb R}^n$. Set $T_{\rm {max}}$, $T_{\rm A}$ and $T_{\rm B}$.
\FOR {$t=1,\cdots,T_{\rm {max}}$ }
\FOR {$t_{\rm B}=1,\cdots,T_{\rm B}$ }
\STATE //\quad [Step 1 (B)] Perform output linear step
\STATE  Calculate $\hat{\mathbf p}_{{\rm B}}(t)$ (\ref{hatpBvec}) and ${\boldsymbol \tau}^p_{{\rm B}}(t)$  (\ref{taupBvec}).
\STATE //\quad [Step 2 (B)] Perform output linear step
\STATE  Calculate $\hat{s}_{{\rm B},i}(t)$ (\ref{hatsB}) and ${\tau}_{{\rm B},i}(t)$ (\ref{tauB}).
\STATE //\quad [Step 3 (B)] Perform input linear step
\STATE  Calculate the posterior means $\hat{\mathbf r}_{{\rm B}}(t)$ (\ref{hatrB}) and variances ${\boldsymbol\tau}^r_{{\rm B}}(t)$ (\ref{taurB}) of $\mathbf{b}$.
\STATE //\quad [Step 4 (B)] Perform input nonlinear step
\STATE  Calculate the posterior means (\ref{hatb}) and variances (\ref{taub}) of $\mathbf{b}$.
\ENDFOR
\STATE Set ${\mathbf b}_{{\rm B}}^{\rm ext}(t)=\hat{\mathbf r}_{{\rm B}}(t)$ and ${\mathbf v}_{{\rm B},b}^{{\rm ext}}(t)={\boldsymbol\tau}^r_{{\rm B}}(t)$.
\FOR {$t_{\rm A}=1,\cdots,T_{\rm A}$ }
\STATE //\quad [Step 1 (A)] Perform output nonlinear step.
\STATE Calculate $\hat{\mathbf s}_{{\rm A}}(t)\in {\mathbb R}^m$ (\ref{hatsA}) and ${\boldsymbol \tau}^s_{{\rm A}}(t) \in {\mathbb R}^m$ (\ref{hattausA}).
\STATE ${\mathbf z}_{\text {B}}^{\text {post}}(t)={\text E}[{\mathbf z}|{\mathbf y},{\mathbf z}_{\text A}^{\text {ext}}(t),{\mathbf v}_{\text A}^{\text {ext}}(t)]$
\STATE //\quad [Step 2 (A)] Perform input linear step
\STATE Calculate $\hat{\mathbf r}_{{\rm A}}(t)\in {\mathbb R}^m$ (\ref{hatrA}) and ${\boldsymbol \tau}_{{\rm A}}^r(t)\in {\mathbb R}^m$ (\ref{taurA}).
\STATE //\quad [Step 3 (A)] Perform input nonlinear step
\STATE Calculate the posterior means (\ref{hatx}) and variances (\ref{taux}) of $\mathbf x$,
\STATE //\quad [Step 4 (A)] Perform output linear step
\STATE Calculate ${\boldsymbol \tau}^p_{{\rm A}}(t)$ (\ref{hatx}) and variances $\hat{\mathbf p}_{{\rm A}}(t)$ (\ref{hatpA})
\ENDFOR
\STATE Set ${\mathbf b}_{{\rm A}}^{\rm ext}(t)=\hat{\mathbf p}_{{\rm A}}(t)$ and ${\mathbf v}_{{\rm A}}^{{\rm ext},b}(t)={\boldsymbol \tau}^p_{{\rm A}}(t)$.
\ENDFOR
\STATE Return $\hat{\mathbf x}$.
\end{algorithmic}
\end{algorithm}
As shown in \cite{meng2}, the extrinsic means ${\mathbf b}_{{\rm B}}^{\rm ext}(t)$ and variances ${\mathbf v}_{{\rm B}}^{{\rm ext},b}(t)$ can be regarded as the pseudo observations and variances of $\mathbf b$ in module A, i.e.,
\begin{align}
	\tilde{\mathbf b}(t)={\mathbf Q}{\mathbf x}+\tilde{\boldsymbol\xi}(t), \label{pseudomodelA}
\end{align}
where $\tilde{\mathbf b}(t)\triangleq {\mathbf b}_{{\rm B}}^{\rm ext}(t)$, $\tilde{\boldsymbol\xi}(t)\sim \mathcal{N}({\mathbf 0},{\rm diag}({\mathbf v}_{{\rm B}}^{{\rm ext},b}(t)))$. Consequently,  we could run standard AMP with $T_{\rm A}$ iterations on this pseudo-linear model (\ref{pseudomodelA}), as shown below.
\vspace{-0.5em}
\begin{itemize}
  \item\quad[Step 1 (A)] For $i=1,2,\cdots,r$, $\hat{s}_{{\rm A},i}(t)$ and $\tau^s_{{\rm A},i}(t)$ can be calculated as
\begin{subequations}
	\begin{align}
		\hat{s}_{{\rm A},i}(t)=\frac{b_{{\rm B},i}^{\rm ext}(t)-\hat p_{{\rm A},i}(t-1)}{v^{{\rm ext},b}_{{\rm B},i}(t)+\tau_{{\rm A},i}^p(t-1)},\label{hatsA}\\
		\tau^s_{{\rm A},i}(t)=\frac{1}{v_{{\rm B},i}^{{\rm ext},b}(t)+\tau^p_{{\rm A},i}(t-1)}.\label{hattausA}
	\end{align}
\end{subequations}
  \item\quad[Step 2 (A)] Perform input linear step to obtain $\hat{\mathbf r}_{{\rm A}}(t)\in{\mathbb R}^n$ and ${\boldsymbol\tau}_{{\rm A}}^r(t)\in{\mathbb R}^n$ as
\begin{subequations}
	\begin{align}
			\boldsymbol{\tau}^r_{\rm A}(t)&=\left(|\mathbf{Q}^{\rm T}|^2\boldsymbol{\tau}^s_{\rm A}(t)\right)^{-1},\label{taurA}\\
			\hat{\mathbf{r}}_{\rm A}(t)&=\hat{\mathbf{x}}(t-1)+\boldsymbol{\tau}^r_{\rm A}(t) \odot \left(\mathbf{Q}^{\rm T}\hat{\mathbf{s}}_{\rm A}(t-1)\right), \label{hatrA}
	\end{align}
\end{subequations}
which can be characterized by a pseudo model
\begin{align}\label{rxmoduleA}
	\hat{\boldsymbol{r}}_{\rm A}(t)=\mathbf{x}+\tilde{\mathbf{w}}(t),
\end{align}
where $\tilde{\mathbf{w}}(t)\sim \mathcal{N}(\mathbf{0},{\rm diag}(\boldsymbol{\tau}^{r}_{\rm A}(t)))$.
  \item\quad[Step 3 (A)] Perform input nonlinear step in Module A to obtain the posterior means and of variances of $\mathbf x$.
\begin{subequations}
	\begin{align}
\hat{x}_j(t)&={\rm E}[x_{j}|\hat{r}_{{\rm A},j}(t),\tau^r_{{\rm A},j}(t)],\label{hatx}\\
\tau^x_j(t)&={\rm Var}[x_{j}|\hat{r}_{{\rm A},j}(t),\tau^r_{{\rm A},j}(t)]\label{taux}
	\end{align}
\end{subequations}
for $j=1,2,\cdots,n$, where ${\rm E}[\cdot]$ and ${\rm Var}[\cdot]$ denotes the posterior means and variances with respect to the likelihood calculated in (\ref{rxmoduleA}) and the prior $p({\mathbf x})$. Even if the nuisance parameters in $p({\mathbf x})$ are unknown, EM algorithm \cite{dempster1977maximum} can be easily incorporated into the GUAMP to learn them similarly as \cite{PhilipEM}.
\vspace{-0.5em}
  \item\quad[Step 4 (A)] Perform output linear step to obtain ${\boldsymbol \tau}^p_{{\rm A}}(t)\in{\mathbb R}^r$ and $\hat{\mathbf p}_{{\rm A}}(t)\in{\mathbb R}^r$, i.e.,
\begin{subequations}\label{simcal}
\begin{align}
{\boldsymbol \tau}^p_{{\rm A}}(t)&=|{\mathbf Q}|^2{\boldsymbol \tau}^x(t), \label{taupA}\\
\hat{\mathbf p}_{{\rm A}}(t)&={\mathbf  Q}\hat{\mathbf x}(t)-{\boldsymbol \tau}^p_{\rm A}(t)\odot\hat{\mathbf s}_{{\rm A}}(t-1).\label{hatpA}
\end{align}
\end{subequations}
\end{itemize}
After running $T_{\rm A}$ iterations of AMP on module A, according to \cite{meng2}, the extrinsic mean ${\mathbf b}_{{\rm A}}^{\rm ext}(t)$ and variance ${\mathbf v}_{{\rm A}}^{{\rm ext},b}(t)$ are
\begin{align}
{\mathbf b}_{{\rm A}}^{\rm ext}(t)=\hat{\mathbf p}_{{\rm A}}(t),
{\mathbf v}_{{\rm A}}^{{\rm ext},b}(t)={\boldsymbol \tau}^p_{{\rm A}}(t).
\end{align}
Now we could again run the GAMP  on module B.  The pseudo code of GUAMP is  summarized as Algorithm \ref{GUAMP}.
\section{Numerical Simulation}
\begin{figure}[t!]
		\centering
		\includegraphics[width=7cm, height=5cm]{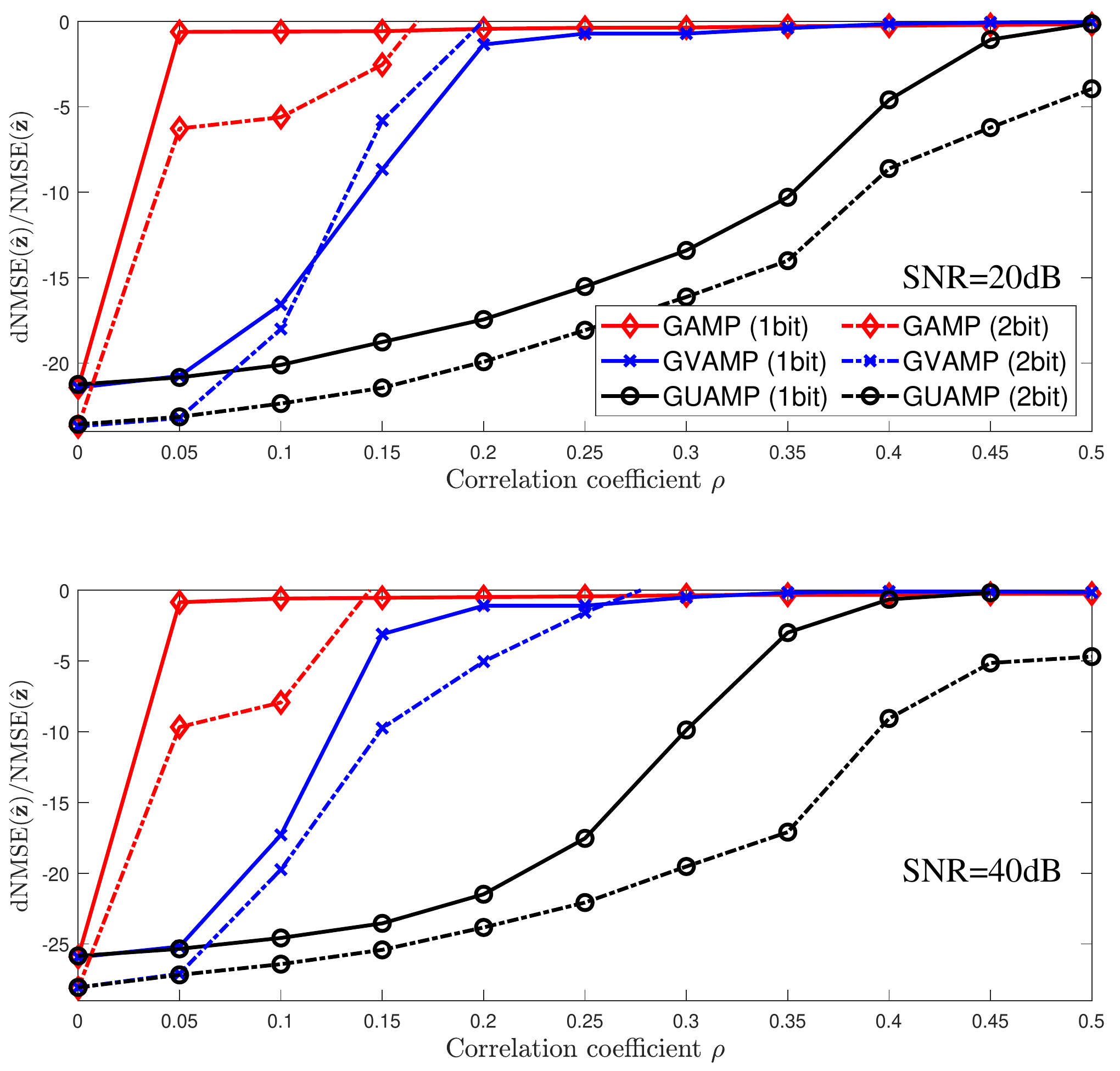}
		\caption{The ${\rm dNMSE}(\hat{\mathbf z})$ ($1$ bit) ${\rm NMSE}(\hat{\mathbf z})$ ($2$ bit) and versus the correlation coefficient $\rho$. (a): ${\rm SNR}=20$ dB, (b): ${\rm SNR}=40$ dB.}
		\label{dNMSEvsRho}
\end{figure}
\begin{figure}[t!]
		\centering
		\includegraphics[width=7cm, height=5cm]{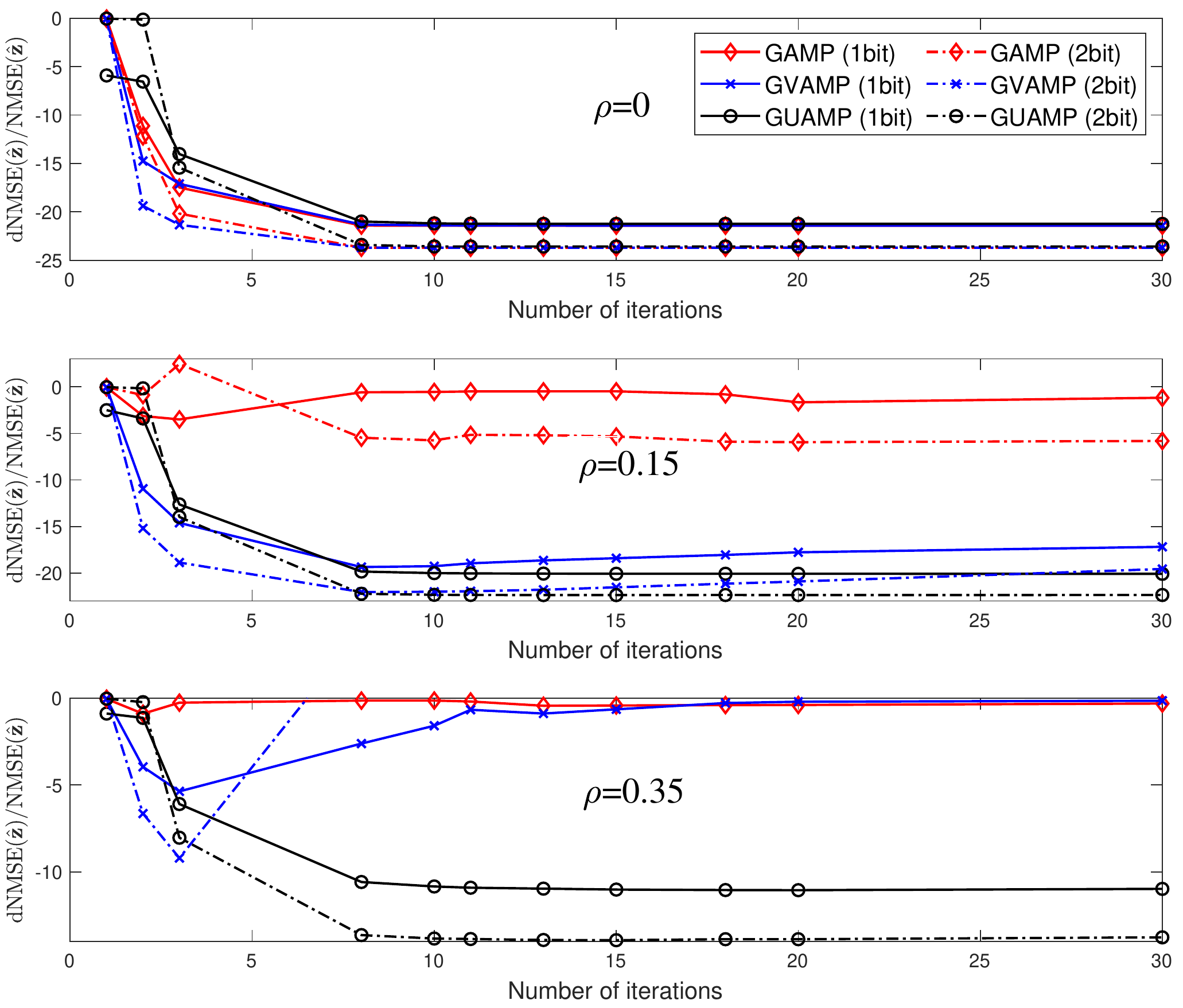}
		\caption{The ${\rm dNMSE}(\hat{\mathbf z})$ ($1$ bit) and ${\rm NMSE}(\hat{\mathbf z})$ ($2$ bit) versus the the number of iterations at ${\rm SNR}=20$ dB.}
		\label{dNMSEvsIte}
\end{figure}
\vspace{-1em}
We verify the efficacy of GUAMP on quantized compressed sensing problem with correlated measurement matrices $\bf{A}$. Specifically, consider the correlated case where ${\mathbf A}$ is constructed as ${\mathbf A}={\mathbf R}_L{\mathbf H}{\mathbf R}_{R}$ \cite{shiu2000fading}, where ${\mathbf R}_L={\mathbf R}_1^{\frac{1}{2}}\in{\mathbb R}^{m\times m}$ and ${\mathbf R}_U={\mathbf R}_2^{\frac{1}{2}}\in{\mathbb R}^{n\times n}$, the $(i,j)$th element of both ${\mathbf R}_1$ and ${\mathbf R}_2$ is $\rho^{|i-j|}$ and $\rho$ is termed as the correlation coefficient here, ${\mathbf H}\in{\mathbb R}^{m\times n}$ is a random matrix whose elements are drawn i.i.d. from ${\mathcal N}(0,1)$. The elements $x_i$ of ${\mathbf x}$ are drawn from a Bernoulli Gaussian distribution $p(x_i)=(1-\lambda)\delta(x_i)+\lambda{\mathcal N}(x_i;0,1/\lambda)$ and $\lambda=0.1$. The $1$ bit (or probit) model ${\mathbf y}={\rm sign}({\mathbf z}+{\mathbf w})$ and $2$ bit quantization models are considered where ${\mathbf z}={\mathbf A}{\mathbf x}$ and ${\mathbf w}\sim {\mathcal N}({\mathbf 0},\sigma^2{\mathbf I}_m)$. The number of measurements and the number of unknowns are $m=2048$ and $n=512$. For $2$ bit quantization, the dynamic range of the quantizer is restricted to $[-3\sigma_z,3\sigma_z]$, where $\sigma_z^2$ is the variance of $z$ and is calculated to be $\sigma_z^2=\|{\mathbf A}\|_{\rm F}/m=n/m=0.25$ as we normalize the measurement matrix $\mathbf A$ such that $\|{\mathbf A}\|_{\rm F}=n$, and the thresholds for $2$ bit quantization are $\{-1.5,0,1.5\}$. The SNR is defined as ${\rm SNR}=\|{\mathbf A}{\mathbf x}\|_2^2/(m\sigma^2)$. The debiased normalized mean squared error (dNMSE) ${{\rm dNMSE}}(\hat{\mathbf z})\triangleq \underset{c}{\operatorname{min}}~{\|{\mathbf z}-c\hat{\mathbf z}\|_2^2}/{\|{\mathbf z}\|_2^2}$ and NMSE ${{\rm NMSE}}(\hat{\mathbf z})\triangleq {\|{\mathbf z}-\hat{\mathbf z}\|_2^2}/{\|{\mathbf z}\|_2^2}$ are used for $1$ and $2$ bit quantization, respectively. The GUAMP is compared with the GAMP and GVAMP. We set $T_{\rm B}=1$ and $T_{\rm A}=4$ (it is numerically found  $T_{\rm A}>1$ improves the robustness of GUAMP). All the results are averaged over $500$ Monte Carlo (MC) trials. Our code is available at  \url{https://github.com/RiverZhu/GUAMP}. 

Fig. \ref{dNMSEvsRho} shows the average dNMSE ($1$ bit) and NMSE ($2$ bit) versus the correlation coefficient $\rho$. It can be seen that when $\rho$ is small and near zero, GUAMP performs the same as GVAMP, and GAMP. As $\rho$ increases, both GVAMP and GAMP tend to diverge for moderate $\rho>0.05$. By contrast, the proposed GUAMP  achieves significantly better performances for large values of $\rho$.

Fig. \ref{dNMSEvsIte} plots the average dNMSE ($1$ bit) and NMSE ($2$ bit) versus iteration for  different correlation coefficients $\rho\in \{0,0.15,0.35\}$ at ${\rm SNR}=20$ dB. It can be seen that  at $\rho=0$,  all the algorithms accurately recover $\mathbf x$ and converge in about tens of iterations. For $\rho=0.15$, GAMP fails and GVAMP starts to diverge, while at a larger value of $\rho=0.35$, both GAMP and GVAMP fail completely while GUAMP converges remarkably well, showing that  GUAMP is more robust than both GVAMP and GAMP under  correlated matrices.
\vspace{-1em}
\section{Conclusion}
\vspace{-1em}
In this paper, inspired from the UAMP \cite{Guo} and a unified inference framework \cite{meng2}, we  propose a novel algorithm called generalized UAMP (GUAMP) to address the GLM inference problem under general measurement matrices. 
Numerical experiments demonstrate that GUAMP significantly outperforms GAMP and GVAMP under correlated measurement matrices. In current implementation of GUAMP, no damping is used although it is believed to further improve its robustness.  The adaptive step scheme \cite{RanganFixed} is also expected to improve GUAMP's convergence and thus it is interesting to formulate the cost function and unveil the relations between GUAMP and the alternating direction method of multipliers (ADMM), which is left as an important future work.
\newpage

\bibliography{mybib}
\bibliographystyle{ieeetr}

\end{document}